\documentclass[preprint,showpacs,amsmath,amssymb,prc]{revtex4}
\usepackage{mathrsfs}
\usepackage{graphicx,color}
\usepackage{multirow}
\usepackage{dcolumn}
\usepackage{bm}
\usepackage{CJK}

\begin{document}
\begin{CJK*}{GBK}{song}
\title{Self-consistent relativistic quasiparticle random-phase approximation and its applications to charge-exchange excitations and $\beta$-decay half-lives}

\author{Z. M. Niu$^{1}$}
\author{Y. F. Niu$^{2}$}
\author{H. Z. Liang$^{3,4}$}\email{haozhao.liang@riken.jp}
\author{W. H. Long$^5$}
\author{J. Meng$^{6,7,8}$}\email{mengj@pku.edu.cn}
\affiliation{$^1$School of Physics and Material Science, Anhui University,
             Hefei 230039, China}
\affiliation{$^2$INFN, Sezione di Milano, I-20133 Milano, Italy}
\affiliation{$^3$RIKEN Nishina Center, Wako 351-0198, Japan}
\affiliation{$^4$Department of Physics, Graduate School of Science, The University of Tokyo, Tokyo 113-0033, Japan}
\affiliation{$^5$School of Nuclear Science and Technology, Lanzhou University,
             Lanzhou 730000, China}
\affiliation{$^6$State Key Laboratory of Nuclear Physics and Technology, School of Physics, Peking
             University, Beijing 100871, China}
\affiliation{$^7$School of Physics and Nuclear Energy Engineering, Beihang University,
             Beijing 100191, China}
\affiliation{$^8$Department of Physics, University of Stellenbosch,
             Stellenbosch 7602, South Africa}

\date{\today}

\begin{abstract}
The self-consistent quasiparticle random-phase approximation (QRPA) approach is formulated in the canonical single-nucleon basis of the relativistic Hatree-Fock-Bogoliubov (RHFB) theory. This approach is applied to study the isobaric analog states (IAS) and Gamov-Teller resonances (GTR) by taking Sn isotopes as examples. It is found that self-consistent treatment of the particle-particle residual interaction is essential to concentrate the IAS in a single peak for open-shell nuclei and the Coulomb exchange term is very important to predict the IAS energies. For the GTR, the isovector pairing can increase the calculated GTR energy, while the isoscalar pairing has an important influence on the low-lying tail of the GT transition. Furthermore, the QRPA approach is employed to predict nuclear $\beta$-decay half-lives. With an isospin-dependent pairing interaction in the isoscalar channel, the RHFB+QRPA approach almost completely reproduces the experimental $\beta$-decay half-lives for nuclei up to the Sn isotopes with half-lives smaller than one second. Large discrepancies are found for the Ni, Zn, and Ge isotopes with neutron number smaller than $50$, as well as the Sn isotopes with neutron number smaller than $82$. The potential reasons for these discrepancies are discussed in detail.
\end{abstract}


\maketitle

\section{Introduction}
Exotic nuclei far from the $\beta$-stability line have become an active field of research, as lots of Radioactive-Ion-Beam (RIB) facilities are operating, being upgraded, under construction, or planned to be constructed \cite{Tanihata1985PRL, Meng1996PRL, Ozawa2000PRL, Vretenar2005PRp, Meng2006PPNP, Paar2007RPP, Otsuka2010PRL,Meng2015JPG}. The charge-exchange excitations of these nuclei play important roles in nuclear physics and various other branches of physics, notably astrophysics. The charge-exchange excitations provide an important probe for studying the spin and isospin properties of the in-medium nuclear interaction. The neutron skin thickness, a basic and critical quantity in nuclear structure, can also be extracted from the sum-rule strengths of the spin-dipole excitations \cite{Krasznahorkay1999PRL}. Moreover, the isobaric analog states (IAS) can be used to study the isospin corrections for the superallowed $\beta$ decays \cite{Hardy2015PRC, Liang2009PRC} and hence to test unitarity of the Cabibbo-Kobayashi-Maskawa matrix. Furthermore, the properties of charge-exchange excitations are essential to predict many nuclear inputs of astrophysics, such as the nuclear $\beta$-decay half-lives, neutrino-nucleus cross sections, and electron-capture cross sections \cite{Langanke2003RMP, Engel1999PRC, Paar2008PRC, Niu2011PRC}. Therefore, nuclear charge-exchange excitations have become one of the hottest topics in nuclear physics and astrophysics.

The charge-exchange excitations can be explored with the charge-exchange reactions, such as $(p, n)$ or $(^3{\rm He}, t)$ reactions, and the weak-decay processes, such as $\beta$ decays \cite{Osterfeld1992RMP, Fujita2011PPNP, Frekers2013NPA}. Although the measurement of charge-exchange excitations has achieved great progress in recent years, their theoretical studies are still essential to understand the microscopic mechanism and also indispensable to many astrophysical applications. Two types of microscopic approaches are widely used in the theoretical investigations on the charge-exchange excitations, the shell model and the quasiparticle random phase approximation (QRPA) approach. Due to the limitation of large configuration space, the shell model calculations are still not feasible for the heavy nuclei away from the magic numbers \cite{Langanke2003RMP, Caurier2005RMP, Koonin1997PRp, Pinedo1999PRL, Garcia2007EPJA, Zhi2013PRC}. However, the QRPA approach can be applied to all nuclei except a few very light systems.

The QRPA approach can be formulated based on the mean-field basis predicted with the empirical potential, such as the deformed Nilsson model \cite{Krumlinde1984NPA, Staudt1990ADNDT, Hirsch1992ADNDT, Nabi2012arXiv}, the finite-range droplet model with a folded Yukawa single-particle potential \cite{Moller1990NPA, Moller1997ADNDT, Moller2003PRC}, and Woods-Saxon potential \cite{Hektor2000PRC, Ni2012JPG}. In addition, based on the Skyrme Hatree-Fock (HF) model, the RPA calculations have been developed for the charge-exchange excitations $30$ years ago \cite{Auerbach1981PLB, Auerbach1984PRC} and has been extended to the QRPA approach by including the pairing correlations for better describing the charge-exchange excitations of open-shell nuclei \cite{Sarriguren2010PRC, Sarriguren2014PRC}. However, the residual interactions used in these QRPA approaches are not directly derived from the interactions used to obtain the mean-field basis. Recently, the self-consistent QRPA approach has received more and more attention, since it is usually believed to possess a better ability of extrapolation. The self-consistent QRPA approaches have been developed based on the Skyrme HF+BCS model \cite{Fracasso2005PRC, Fracasso2007PRC} and Skyrme Hatree-Fock-Bogoliubov (HFB) model \cite{Engel1999PRC, Li2008PRC}. Moreover, the important ingredient of nuclear force --- the tensor force was found to play a crucial role in describing the nuclear charge-exchange excitations and $\beta$-decay half-lives within the QRPA approaches~\cite{Bai2009PLB, Bai2010PRL, Minato2013PRL, Bai2013PLB}, which inspires much interest to explore the nature of nuclear tensor force~\cite{Jiang2015-1, Jiang2015-2}.

During the past years, the covariant density functional theory has successfully described many nuclear phenomena \cite{Ring1996PPNP, Vretenar2005PRp, Meng2006PPNP, Paar2007RPP, Niksic2011PPNP, Liang2015PRp, Meng2015JPG, Meng2016Book} and their predictions are also successfully applied to the simulations of rapid neutron-capture process ($r$ process) \cite{Sun2008PRC, Niu2009PRC, Xu2013PRC, Niu2013PLB}. The self-consistent RPA approach was first developed based on the relativistic Hatree (RH) model \cite{Conti1998PLB}. The negative-energy states in the Dirac sea are found to be very important to construct the RPA configuration space, which remarkably influence the isoscalar strength distributions \cite{Ring2001NPA} and the sum rule of Gamow-Teller (GT) transitions \cite{Ma2004EPJA}. Furthermore, the QRPA approach is formulated in the canonical single-nucleon basis of the relativistic Hartree-Bogoliubov (RHB) theory and used to study nuclear multipole excitations of open-shell nuclei \cite{Paar2003PRC}. The RHB+QRPA approach is then extended to study nuclear charge-exchange excitations \cite{Paar2004PRC, Finelli2007NPA} and further to calculate $\beta$-decay half-lives not only for neutron-rich nuclei \cite{Niksic2005PRC, Marketin2007PRC, Wang2016JPG} but also for the neutron-deficient nuclei \cite{Niu2013PRCR}. Recently, a systematic calculation on nuclear $\beta$-decay properties, including half-lives, $\beta$-delayed neutron emission probabilities, and the average number of emitted neutrons, was performed with the RHB+QRPA model for $5409$ nuclei in the neutron-rich region of the nuclear chart \cite{Marketin2016PRC}.

For the QRPA approaches in the relativistic Hartree approximation, the isovector $\pi$ meson plays an important role in the description of nuclear charge-exchange resonances, while this degree of freedom is absent in the ground-state description due to the parity conservation. To account for the contact interaction coming from the pseudovector pion-nucleon coupling, a zero-range counter term is introduced, while its strength is treated as an adjustable parameter to reproduce experimental data on the GT excitation energies. In the relativistic HF (RHF) approximation, the contributions of $\pi$ meson to the nuclear ground-state properties can be naturally included via the exchange (Fock) terms and the description of the nucleon effective mass and the nuclear shell structures is improved \cite{Long2006PLB, Long2007PRC}. Based on the RHF model, the fully self-consistent relativistic RPA (RHF+RPA) approach has been developed. The RHF+RPA model achieves an excellent agreement on the data of Gamow-Teller resonances (GTR) and spin-dipole resonances (SDR) in doubly magic nuclei, without any readjustment of the parameters of the covariant energy density functional including the zero-range counter term \cite{Liang2008PRL, Liang2012PRC}.

To provide an accurate and reliable description of open-shell nuclei, the pairing correlations have to be treated in proper way. By combining with the BCS method, the RHF+BCS model has been formulated and it is found that the description of nuclear shell evolution along isotopic chain of $Z=50$ and isotonic chain of $N=82$ can be improved with the presence of the degree of freedom associated with the pion pseudovector coupling~\cite{Long2008EPL, Long2009PLB}. Extending to the neutron/proton drip line, the pairing gap energy becomes comparable to the nucleon separation energy and the continuum effects can be involved substantially by the pairing correlation. It thus requires a unified description of mean field and pairing correlations, for instance within the Bogoliubov scheme~\cite{Dobaczewski1984NPA, Meng1998NPA, Meng2006PPNP}. Integrated with the Bogoliubov transformation, the relativistic Hartree-Fock-Bogoliubov (RHFB) theory was developed recently~\cite{Long2010PRC} and it achieved great success in the description of the exotic nuclei far from the $\beta$-stability line~\cite{Long2010PRCR, Wang2013PRC-1, Wang2013PRC-2, Lu2013PRC, Li2015PRC, Li2016PLB} and superheavy nuclei \cite{Li2014PLB}. Based on the RHFB theory, the self-consistent QRPA (RHFB+QRPA) approach was developed and a systematic study on the $\beta$-decay half-lives of neutron-rich even-even nuclei with $20\leqslant Z\leqslant 50$ has been performed \cite{Niu2013PLB}.

In this work, we will employ the RHFB+QRPA approach to investigate the charge-exchange excitations, including the IAS and GTR. Furthermore, the nuclear $\beta$-decay half-lives predicted with the RHFB+QRPA approach will be presented and compared with the experimental data and other theoretical results. These results are given in Sec.~\ref{Sec:3}. In Sec.~\ref{Sec:2}, the basic formulas of RHFB theory, QRPA approach, and the calculations of nuclear $\beta$-decay half-lives are briefly introduced. Finally, summary and perspectives are presented in Sec.~\ref{Sec:4}.

\section{Theoretical framework}\label{Sec:2}
In this Section, the basic formulas of the RHFB theory will be briefly introduced, then the self-consistent QRPA approach based on the RHFB theory will be formulated in the canonical basis of the RHFB framework. With the transition properties obtained from the QRPA approach, the calculations of nuclear $\beta$-decay half-lives will be also presented.

\subsection{Effective Lagrangian density}
The basic ansatz of the RHF theory is a Lagrangian density where nucleons are described as Dirac particles which interact to each other via the exchange of mesons ($\sigma$, $\omega$, $\rho$, and $\pi$) and the photon ($A$),
\begin{eqnarray}\label{Eq:LagDen}
  {\cal L} &=& \bar{\psi} \left[   i\gamma^\mu\partial_\mu - M - g_\sigma\sigma
                                - \gamma^\mu g_\omega\omega_\mu + g_\rho\gamma^\mu\vec{\tau}\cdot\vec{\rho}_\mu \right. \nonumber\\
          & &            \left. - \frac{f_\pi}{m_\pi}\gamma_5\gamma^\mu\partial_\mu\vec{\pi}\cdot\vec{\tau}
                                + e\gamma^\mu\frac{1-\tau_3}{2}A_\mu
                         \right]\psi \nonumber\\
          & &+\frac{1}{2}\partial^\mu\sigma\partial_\mu\sigma - \frac{1}{2}m_\sigma^2\sigma^2
             -\frac{1}{4}\Omega^{\mu\nu}\Omega_{\mu\nu} + \frac{1}{2}m_\omega^2\omega^\mu\omega_\mu \nonumber\\
          & &-\frac{1}{4}\vec{R}^{\mu\nu}\cdot\vec{R}_{\mu\nu} + \frac{1}{2}m_\rho^2\vec{\rho}^\mu\cdot\vec{\rho}_\mu
             +\frac{1}{2}\partial^\mu\vec{\pi}\cdot\partial_\mu\vec{\pi} \nonumber\\
          & &-\frac{1}{2}m_\pi^2\vec{\pi}\cdot\vec{\pi}-\frac{1}{4}F^{\mu\nu}F_{\mu\nu},
\end{eqnarray}
where $M$ and $m_i$ ($i=\sigma, \omega, \rho$, and $\pi$) are the masses of the nucleon and mesons, $g_\sigma, g_\omega, g_\rho$, and $f_\pi$ are meson-nucleon couplings, respectively. The field tensors for the vector mesons and the photon are defined as
\begin{eqnarray}
  \Omega_{\mu\nu}  &=& \partial_\mu\omega_\nu - \partial_\nu\omega_\mu, \nonumber\\
  \vec{R}^{\mu\nu} &=& \partial_\mu\vec{\rho}_\nu - \partial_\nu\vec{\rho}_\mu, \nonumber\\
  F_{\mu\nu}       &=& \partial_\mu A_\nu - \partial_\nu A_\mu.
\end{eqnarray}
Following the standard variational procedure of the Lagrangian density, one can obtain the Euler-Lagrange canonical field equations, which just correspond to the Dirac, Klein-Gordon, and Proca equations for the nucleon, meson, and photon fields, respectively. As these equations are too difficult to be solved exactly, one has to introduce some reasonable approximations, such as the Hartree or Hartree-Fock approximations.

\subsection{Energy functional and Dirac Hartree-Fock equation}\label{Sec:EFandDHF}

Before applying the Hartree or Hartree-Fock approximations, the energy functional should be firstly built up by taking the expectation value of Hamiltonian. The Hamiltonian density can be obtained with the general Legendre transformation,
\begin{eqnarray}
  {\cal H}=\frac{\partial{\cal L}}{\partial{\dot{\phi}_i}}\dot{\phi}_i-{\cal L},
\end{eqnarray}
where $\phi_i$ represents the nucleon field $\psi$, the $\sigma$-, $\omega$-, $\rho$-, and $\pi$-meson fields, and the photon field $A$. Combing the field equations of mesons and photon, the Hamiltonian $H =\int \mathrm{d}^3 x {\cal H}$ in the nucleon space can be expressed as
\begin{eqnarray}\label{Eq:Hamilt}
  H &=&\int \mathrm{d}^3 x_1 \bar{\psi}(-i\boldsymbol{\gamma}\cdot\boldsymbol{\nabla}+M)\psi
       +\frac12\iint \mathrm{d}^3 x_1 \mathrm{d}^4 x_2\ \nonumber\\
    & &            \sum_{\begin{subarray}{l}
                         i=\sigma,\omega,\\
                         \rho,\pi,A\\
                        \end{subarray}}
                   \bar{\psi}(x_1)\bar{\psi}(x_2)
                   \Gamma_i(1,2) D_i(1,2)
                   \psi(x_2)\psi(x_1),~~~~
\end{eqnarray}
where the two-body interaction vertices $\Gamma_i(1,2)$ for the meson and photon fields are
\begin{eqnarray}
  \Gamma_\sigma(1,2) &=& - g_\sigma(1)g_\sigma(2),\\
  \Gamma_\omega(1,2) &=& + g_\omega(1)\gamma_\mu(1)g_\omega(2)\gamma^\mu(2),\\
  \Gamma_\rho(1,2)   &=& + g_\rho(1)\gamma_\mu(1)\vec{\tau}(1)
                           \cdot
                           g_\rho(2)\gamma^\mu(2)\vec{\tau}(2),\\
  \Gamma_\pi(1,2)    &=& - \Big[\frac{f_\pi}{m_\pi}\vec{\tau}\gamma_5\gamma_\mu\partial^\mu\Big]_1
                           \cdot
                           \Big[\frac{f_\pi}{m_\pi}\vec{\tau}\gamma_5\gamma_\nu\partial^\nu\Big]_2,\\
  \Gamma_A(1,2)      &=& + \frac{e^2}{4}\big[\gamma_\mu(1-\tau_3)\big]_1\big[\gamma^\mu(1-\tau_3)\big]_2.
\end{eqnarray}
Neglecting the retardation effects, the propagators $D_i(1,2)$ for the meson and photon fields can be simplified to be
\begin{eqnarray}
    D_{i}(1,2)
  &=& \frac{1}{4\pi}\frac{e^{m_i|x_1-x_2|}}{|x_1-x_2|},\\
     D_A(1,2)
  &=& \frac{1}{4\pi}\frac{1}{|x_1-x_2|}.
\end{eqnarray}

To quantize the Hamiltonian $H$ in Eq.~(\ref{Eq:Hamilt}), the nucleon field operators $\psi$ and $\bar{\psi}$ are expanded on the set of creation and annihilation operators of nucleons $(c_\alpha^\dag, c_\alpha)$ and antinucleons $(d_\alpha^\dag, d_\alpha)$,
\begin{eqnarray}
   \psi(x)     &=&\sum_\alpha \left[ f_\alpha(\boldsymbol{x}) e^{-i\varepsilon_\alpha t} c_\alpha
                                    +g_\alpha(\boldsymbol{x}) e^{-i\varepsilon'_\alpha t} d_\alpha^\dag \right],\\
   \psi^\dag(x)&=&\sum_\alpha \left[ f_\alpha^\dag(\boldsymbol{x}) e^{-i\varepsilon_\alpha t} c_\alpha^\dag
                                    +g_\alpha^\dag(\boldsymbol{x}) e^{-i\varepsilon'_\alpha t} d_\alpha \right],
\end{eqnarray}
where $f_\alpha(x)$ and $g_\alpha(x)$ are the Dirac spinors in a state $\alpha$. The inclusion of $d_\alpha$ and $d^\dag_\alpha$ terms leads to divergences and requires a cumbersome renormalization procedure \cite{Serot1986ANP}, so these terms are usually omitted in the expansions, i.e., the so-called no-sea approximation. Then, the Hamiltonian can be expressed as
\begin{eqnarray}\label{Eq:HamiltQuant}
   H = \sum_{\alpha\beta}T_{\alpha\beta}c_\alpha^\dag c_\beta
      +\frac12 \sum_i\sum_{\alpha\alpha'\beta\beta'}
               V_{\alpha\beta\beta'\alpha'}^i
               c_\alpha^\dag c_\beta^\dag c_{\beta'} c_{\alpha'}
\end{eqnarray}
with the kinetic term $T$ and two-body interaction terms $V^i$,
\begin{eqnarray}
        T_{\alpha\beta}
   &=& \int \mathrm{d} \boldsymbol{x}
            \bar{f_\alpha}(i\boldsymbol{\gamma}\cdot\boldsymbol{\nabla}+M)f_\beta, \\
        V_{\alpha\beta\beta'\alpha'}^i
   &=&
       \iint \mathrm{d} \boldsymbol{x_1}\mathrm{d} \boldsymbol{x_2}
            \bar{f_\alpha}(1)\bar{f_\beta}(2) \Gamma_i(1,2) \nonumber\\
   & &      D_i(1,2)f_{\beta'}(2)f_{\alpha'}(1).
\end{eqnarray}

In the Hartree-Fock approximation, the trial ground state is chosen as a Slater determinant, i.e.,
\begin{eqnarray}\label{Eq:Phi0}
   |\Phi_0\rangle = \prod_\alpha c_\alpha^\dag |0\rangle,
\end{eqnarray}
with the vacuum $|0\rangle$. The energy functional is then obtained from the expectation with respect to the ground state $\left|\Phi_0\right>$,
\begin{eqnarray}\label{Eq:Efunc1}
   E&=& \langle \Phi_0| H |\Phi_0\rangle =\left<\Phi_0\right| (T+\sum_i V^i) \left|\Phi_0\right>.
\end{eqnarray}
The expectation of the two-body interaction term $V^i$ will lead to two type of contributions, namely the direct (Hartree) and exchange (Fock) terms. With only the direct term, Eq. (\ref{Eq:Efunc1}) just corresponds to the energy functional of the RMF or RH theory, while with both direct and exchange
terms, one obtains the energy functional of the RHF theory.

Taking the variation of the energy functional (\ref{Eq:Efunc1}) with respect to the Dirac spinor $f_\alpha$, one then gets the Dirac Hartree-Fock equation,
\begin{eqnarray}\label{Eq:DiracEq}
   \int \mathrm{d}\boldsymbol{r}' h(\boldsymbol{r},\boldsymbol{r}')f_\alpha(\boldsymbol{r}')=\varepsilon_\alpha f_\alpha(\boldsymbol{r}),
\end{eqnarray}
where $h(\boldsymbol{r},\boldsymbol{r}')$ is the single-particle Dirac Hamiltonian and $\varepsilon$ is the single-particle energy including the rest mass. There are three parts for $h(\boldsymbol{r},\boldsymbol{r}')$, i.e., $h=h^{\rm kin}+h^{\rm D}+h^{\rm E}$. They respectively denote the kinetic energy, the direct local potential, and the exchange nonlocal potential. The readers can refer to Refs.~\cite{Long2010PRC} for the detailed expressions of $h^{\rm kin}$, $h^{\rm D}$, and $h^{\rm E}$.

\subsection{Relativistic Hartree-Fock-Bogoliubov theory}
To describe the properties of open-shell nuclei, the pairing correlations should be included, which is taken into account with the Bogoliubov theory in this work. Following the standard procedure of the Bogoliubov transformation \cite{Gorkov1958SP, Ring1980Book, Kucharek1991ZPA}, one then obtains the relativistic Hartree-Fock-Bogoliubov equation as
\begin{align}\label{Eq:RHFBEq}
  \int \mathrm{d} \boldsymbol{r}' &\begin{pmatrix}  h(\boldsymbol{r},\boldsymbol{r}') & \Delta(\boldsymbol{r},\boldsymbol{r}') \\[0.5em]
  \Delta(\boldsymbol{r},\boldsymbol{r}') & -h(\boldsymbol{r},\boldsymbol{r}')\end{pmatrix} \begin{pmatrix} f_U(\boldsymbol{r}') \\[0.5em] f_V(\boldsymbol{r}')\end{pmatrix} \nonumber \\&\hspace{5em} = \begin{pmatrix}  E+\lambda & 0 \\[0.5em]  0 & E-\lambda \end{pmatrix}\begin{pmatrix} f_U(\boldsymbol{r}) \\[0.5em] f_V(\boldsymbol{r})\end{pmatrix},
\end{align}
where $f_U$ and $f_V$ are the quasiparticle spinors and $\lambda$ is the chemical potential. The pairing potential $\Delta(\boldsymbol{r},\boldsymbol{r}')$ can be expressed to be
\begin{eqnarray}\label{Eq:PairingPotential}
    \Delta(\boldsymbol{r},\boldsymbol{r}')
   =-\frac12\sum_\beta
                       V_{\alpha\beta}^{pp}(\boldsymbol{r},\boldsymbol{r}')
                       \kappa_\beta(\boldsymbol{r},\boldsymbol{r}'),
\end{eqnarray}
where the pairing tensor is
\begin{eqnarray}\label{Eq:PairingTensor}
    \kappa_\beta(\boldsymbol{r},\boldsymbol{r}')
   =f_{V_\alpha}(\boldsymbol{r})^* f_{U_\alpha}(\boldsymbol{r}').
\end{eqnarray}
For the pairing interaction $V^{pp}$, we adopt the pairing part of the Gogny force
\begin{align}\label{Eq:Gogny}
    V^{pp}(\boldsymbol{r},\boldsymbol{r}')
   =&\sum_{i=1,2}e^{[(\boldsymbol{r}-\boldsymbol{r}')/\mu_i]^2}
                (W_i + B_i P^\sigma - H_i P^\tau\nonumber\\
    &\qquad          - M_i P^\sigma P^\tau),
\end{align}
with the set D1S \cite{Berger1984NPA} for the parameters $\mu_i, W_i, B_i, H_i$, and $M_i$.

In this work, the spherical symmetry is assumed for the nuclear systems and the RHFB equation is solved by an expansion of quasiparticle spinors in the Dirac Woods-Saxon (DWS) basis \cite{Zhou2003PRC, Long2010PRC}. The numbers of positive- and negative-energy states in the DWS basis are taken as $N_F=28$ and $N_D=20$, respectively. Details of solving the RHFB equations in the DWS basis can be found in Ref.~\cite{Long2010PRC}.

\subsection{Quasiparticle random phase approximation}
The QRPA equations can be derived from the time-dependent RHFB theory in the limit of small-amplitude oscillations similar to Refs.~\cite{Paar2003PRC, Paar2004PRC}. Previous studies have found that the QRPA equations can be easily solved in the canonical basis, in which the RHFB wave functions are expressed in the form of BCS-like wave functions. With the spherical symmetry, the quasiparticle pairs can be coupled to a good angular momentum and the matrix equations of the QRPA for the charge-exchange excitations read
\begin{align}\label{Eq:QRPAEq}
  \begin{pmatrix}  A^J_{pnp'n'}    &  B^J_{pnp'n'} \\[0.5em]  -B^{*J}_{pnp'n'} & -A^{*J}_{pnp'n'}\end{pmatrix}\begin{pmatrix}
    X^{\nu J}_{p'n'} \\[0.5em]      Y^{\nu J}_{p'n'}   \end{pmatrix} = & E_\nu \begin{pmatrix}  X^{\nu J}_{pn} \\[0.5em]
       Y^{\nu J}_{pn}\end{pmatrix},
\end{align}%
where $p$, $p'$, and $n$, $n'$ denote proton and neutron quasiparticle canonical states, respectively. For each transition energy $E_\nu$, quantities $X^{\nu J}_{pn}$ and $Y^{\nu J}_{pn}$ denote the corresponding forward- and backward-going QRPA amplitudes, respectively. The angular-momentum coupled matrix elements $A^J$ and $B^J$ read
\begin{eqnarray}
       A^J_{pnp'n'}
    &=&H^{11}_{pp'}\delta_{nn'} + H^{11}_{nn'}\delta_{pp'}\nonumber\\
    &+&H^{phJ}_{pnp'n'}(u_p v_n u_{p'} v_{n'} + v_p u_n v_{p'} u_{n'})\nonumber\\
    &+&H^{ppJ}_{pnp'n'}(u_p u_n u_{p'} u_{n'} + v_p v_n v_{p'} v_{n'}),\\ \label{Eq:QRPAA}
       B^J_{pnp'n'}
    &=&H^{phJ}_{pnp'n'}(u_p v_n v_{p'} u_{n'} + v_p u_n u_{p'} v_{n'})\nonumber\\
    &-&H^{ppJ}_{pnp'n'}(u_p u_n v_{p'} v_{n'} + v_p v_n u_{p'} u_{n'}),\label{Eq:QRPAB}
\end{eqnarray}
with
\begin{eqnarray}\label{Eq:H11}
   H^{11}_{kk'}= h_{kk'}(u_k u_{k'} - v_{k} v_{k'})
                -\Delta_{kk'}(u_k v_{k'} + v_{k} u_{k'}).
\end{eqnarray}
The terms $H^{phJ}$ and $H^{ppJ}$ in matrix elements $A^J$ and $B^J$ denote the contributions from particle-hole (ph) and particle-particle (pp) interactions, respectively.

In the self-consistent QRPA approach based on the RHFB theory, the contributions from exchange terms must be included, so the term $H^{phJ}$ corresponding to the ph interaction $V^{ph}$ is
\begin{eqnarray}
    H^{phJ}_{pnp'n'} = V^{phJ}_{pn'np'} - V^{phJ}_{pn'p'n}.
\end{eqnarray}
In this work, $V^{ph}$ includes the contributions from the $\sigma$-, $\omega$-, $\rho$-, and $\pi$-meson fields, i.e.,
\begin{eqnarray}
    V^{ph}=\sum_{i=\sigma,\omega,\rho,\pi}\Gamma_i(1,2)D_i(1,2),
\end{eqnarray}
where $\Gamma_i(1,2)$ and $D_i(1,2)$ are the interaction vertices and propagators of corresponding meson fields given in Sec.~\ref{Sec:EFandDHF}. In addition, a zero-range pionic counter term should be included to cancel the contact interaction coming from the pion pseudovector coupling, which reads
\begin{align}
  V_{\pi}^{\delta}(1,2) = & - \frac{1}{3}\Big[\frac{f_\pi}{m_\pi} \vec\tau \gamma_5\gamma_i\Big]_1\cdot\Big[\frac{f_\pi}{m_\pi}\vec\tau \gamma_5 \gamma^i\Big]_2\delta(\boldsymbol{r}_1-\boldsymbol{r}_2).
\end{align}

Similarly, the term $H^{ppJ}$ corresponding to the pp interaction $V^{pp}$ is
\begin{eqnarray}
    H^{ppJ}_{pnp'n'}=V^{ppJ}_{pnp'n'}-V^{ppJ}_{pnn'p'}.
\end{eqnarray}
In the isovector ($T=1$) pp channel, we adopt the pairing part of the Gogny force with the parameter set D1S as in the RHFB ground-state calculations. In the isoscalar ($T=0$) pp channel, we employ a finite-range interaction as in Refs.~\cite{Engel1999PRC, Paar2004PRC, Niksic2005PRC, Marketin2007PRC, Niu2013PLB, Niu2013PRCR, Wang2016JPG},
\begin{eqnarray}\label{Eq:VppT0}
    V_{T=0}^{pp}(1,2)=-V_0\sum_{i=1,2}g_i e^{[(\boldsymbol{r}_1-\boldsymbol{r}_2)/\mu_i]^2}
                      \hat{\Pi}_{S=1,T=0},
\end{eqnarray}
with $\mu_1=1.2$~fm, $\mu_2=0.7$~fm, $g_1=1$, and $g_2=-2$. The operator $\hat{\Pi}_{S=1,T=0}$ projects onto states with $S=1$ and $T=0$. For the strength parameter $V_0$, we employ the following ansatz proposed in Ref.~\cite{Niu2013PLB},
\begin{eqnarray}\label{Eq:V0NSubZ}
  V_0 &=& V_L +\frac{V_D}{1+e^{a+b(N-Z)}}\; ,
\end{eqnarray}
with $V_L=134.0$~MeV, $V_D=121.1$~MeV, $a=8.5$, and $b=-0.4$ which provide the best description of available half-life data \cite{Audi2012CPC} in the region $20\leqslant Z\leqslant 50$.

By diagonalizing the QRPA matrix in Eq.~(\ref{Eq:QRPAEq}), one can get the discrete transition energies $E_\nu$ and the corresponding QRPA amplitudes $X^{\nu J}_{pn}$ and $Y^{\nu J}_{pn}$. Then the transition probabilities $B_{\nu J}$ induced by the operator $T^{JM}$ between the ground state of the even-even $(N,Z)$ nucleus and the excited state of the odd-odd $(N+1,Z-1)$ or $(N-1,Z+1)$ nucleus can be calculated by
\begin{eqnarray}
   B_{\nu}=\bigg|\sum_{pn}\langle p \| T^J \| n \rangle
                        [X^{\nu J}_{pn}u_p v_n + (-1)^J Y^{\nu J}_{pn}v_p u_n]
         \bigg|^2.
\end{eqnarray}
The strength distribution is obtained by folding the discrete transition probabilities with Lorentzian function, i.e.,
\begin{eqnarray}
   R(E)=\sum_\nu B_{\nu}
                 \frac{\Gamma/2\pi}{(E-E_\nu)^2+\Gamma^2/4},
\end{eqnarray}
where the width $\Gamma$ is taken to be $1$~MeV for illustrating our calculations of the spin-isospin excitations.

\subsection{Nuclear $\beta$-decay half-lives}
The $\beta$-decay half-life of an even-even nucleus in the allowed GT approximation is calculated with
\begin{eqnarray}\label{Eq:BetaDecayRate}
    T_{1/2}
  =\frac{D}{g_A^2 \sum_{E_D<Q_\beta} B(E_D) f(Z,\omega)},
\end{eqnarray}
where $D=6163.4\pm3.8$~s and $g_A=1$. The $B(E_D)$ is the transition strength from the ground state of mother nucleus to the final state with excitation energy $E_D$, which is referred to the ground state of the daughter nucleus. The summation includes all the final states having an excitation energy $E$ smaller than $Q_\beta$. The integrated phase volume $f(Z,\omega)$ is
\begin{eqnarray}\label{Eq:PhaseVol}
  f(Z,\omega) = 
               \int_{m_e}^{\omega}
               p_e E_e (\omega-E_e)^2 F_0(Z,E_e)\mathrm{d}E_e,
\end{eqnarray}
where $m_e$, $p_e$, $E_e$, and $F_0(Z,E_e)$ denote the rest mass, momentum, energy, and Fermi function of the emitted electron, respectively. The $\beta$-decay transition energy $\omega$, which is the energy difference between the initial and final nuclear states, is calculated by
\begin{equation}\label{Eq:BetaDecayEm1}
    \omega = Q_\beta + m_e - E_D.
\end{equation}

In the present self-consistent QRPA calculation, the excitation energy $E_\nu$ in Eq.~(\ref{Eq:QRPAEq}) is referred to the ground state of the mother nucleus corrected by the mass difference between neutron and proton $\Delta_{np}$ as in Refs.~\cite{Niu2013PLB, Niu2013PRCR, Wang2016JPG}. It is here denoted by $E_M$ to be clearly distinguished from $E_D$. Therefore, one has
\begin{equation}\label{Eq:BetaDecayEQRPA}
    E_M = E_D + \Delta B = E_D + (\Delta_{nH} - Q_\beta),
\end{equation}
where $\Delta B$ is the binding energy difference between mother nucleus and daughter nucleus, and $\Delta_{nH}$ is the mass difference between the neutron and the hydrogen atom. Combining Eqs.~(\ref{Eq:BetaDecayEm1}) and (\ref{Eq:BetaDecayEQRPA}), one obtains
\begin{equation}\label{Eq:BetaDecayEm}
    \omega = \Delta_{nH} + m_e - E_M = \Delta_{np} - E_M.
\end{equation}
Since the energy of the emitted electron must be higher than its rest mass, i.e., $\omega > m_e$, the final nuclear states are those with the excitation energies $E_M < \Delta_{nH}$. Equation~(\ref{Eq:BetaDecayRate}) then becomes
\begin{eqnarray}\label{Eq:BetaDecayRateEM}
    T_{1/2}
  =\frac{D}{g_A^2 \sum_{E_M<\Delta_{nH}} B(E_M) f(Z,\omega)},
\end{eqnarray}
where both $E_M$ and $B(E_M)$ can be directly obtained from the self-consistent QRPA calculations.

\section{Results and discussion}\label{Sec:3}

In the self-consistent QRPA calculations, the reasonable description of nuclear ground-state properties is essential to predict nuclear charge-exchange excitations. Therefore, in this Section, we will first study the description of nuclear ground-state properties by using the RHFB theory. The two-neutron separation energies and the neutron-skin thicknesses will be taken as examples. The self-consistent QRPA calculations based on the RHFB theory will be then shown for the IAS and GTR, on which the effects of the ph and pp residual interactions will be investigated carefully. Finally, the nuclear $\beta$-decay half-lives predicted with the RHFB+QRPA approach will be presented and compared with the experimental data and other theoretical results.
The effective interactions PKO1 \cite{Long2006PLB} and DD-ME2 \cite{Lalazissis2005PRC} are adopted for the RHFB(+QRPA) and RHB(+QRPA) calculations, respectively.

\subsection{Ground-state properties}

\begin{figure}
\includegraphics[width=8cm]{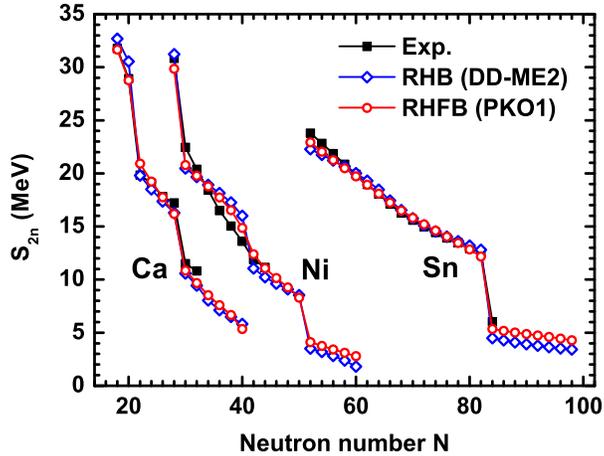}
\caption{(Color online) Two-neutron separation energies of the even-even Ca, Ni, and Sn isotopes.
The RHFB calculations with the effective interaction PKO1 are denoted by the open circles.
For comparison, the experimental data \cite{Wang2012CPC} and the calculated results by the RHB theory with the effective interaction DD-ME2 are shown by the filled squares and open diamonds, respectively.}
\label{Fig:FigS2n}
\end{figure}

Figure~\ref{Fig:FigS2n} shows the two-neutron separation energies of the even-even Ca, Ni, and Sn isotopes calculated by the RHFB theory. It is clear that the RHFB approach well reproduces the experimental data in a rather wide range from $Z=20$ to $Z=50$. It is known that the two-neutron separation energies $S_{2n}$ contain detailed information about the nuclear structure.
The abrupt drop of $S_{2n}$ generally reflects the existence of shell structure. From the abrupt drop of experimental $S_{2n}$ in Fig.~\ref{Fig:FigS2n}, the shell structures at $N=20$, $28$, and $82$ are clearly observed. Both the RHB and RHFB approaches correctly describe the positions of the shell structures. However, the RHB calculations with the effective interaction DD-ME2 overestimate the shell effects at $N=40$ for the Ni isotopes. For the RHFB calculations with the effective interaction PKO1, the strengthes of the shell closures at $N=20$, $28$, and $82$ are satisfactorily reproduced, as well as the shell effects at $N=40$.

\begin{figure}
\includegraphics[width=8cm]{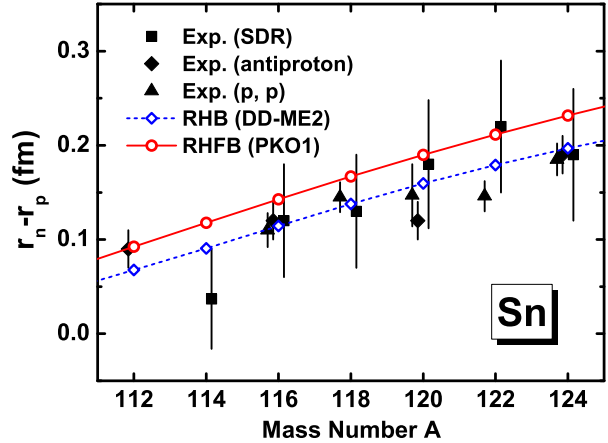}
\caption{(Color online) Neutron-skin thicknesses ($r_n-r_p$) of the even-even Sn isotopes.
Open circles and open diamonds show the results calculated by the RHFB theory with PKO1 and the RHB theory with DD-ME2, respectively. The experimental results from the spin-dipole resonance (SDR) \cite{Krasznahorkay1999PRL}, anti-protonic x-ray data \cite{Trzcinska2001PRL}, and proton elastic scattering \cite{Terashima2008PRC} are shown by the filled squares, diamonds, and triangles, respectively.}
\label{Fig:FigSnRnp}
\end{figure}

The neutron-skin thicknesses of the even-even Sn isotopes are shown in Fig.~\ref{Fig:FigSnRnp}. Generally speaking, the calculations with PKO1 and DD-ME2 reproduce the experimental results from the spin-dipole resonance (SDR) \cite{Krasznahorkay1999PRL}, anti-protonic x-ray data \cite{Trzcinska2001PRL}, and proton elastic scattering \cite{Terashima2008PRC} very well. The exception is the data from SDR for $^{114}$Sn, which
deviates from the systematic trend. Comparing these two approaches, the results of PKO1 are systematically larger than those of DD-ME2. This can be mainly explained by the larger symmetry energy of PKO1, $E_{\rm sym}=34.4$~MeV, in comparison with that of DD-ME2, $E_{\rm sym}=32.3$~MeV, since there exists a linear relation between the neutron-skin thickness and the symmetry energy of nuclear matter at saturation density \cite{Chen2005PRC}. Significant progress has been made on constraining the symmetry energy during the past decades.
Combing the current available constraints on the symmetry energy obtained from terrestrial laboratory measurements and astrophysical observations,  the symmetry energy $E_{\rm sym}=32.5\pm 2.5$~MeV has been concluded \cite{Chen2014NPR}. Obviously, the symmetry energies from both PKO1 and DD-ME2 agree with the constraint.

\subsection{Spin-isospin excitations}

\begin{figure}
\includegraphics[width=8cm]{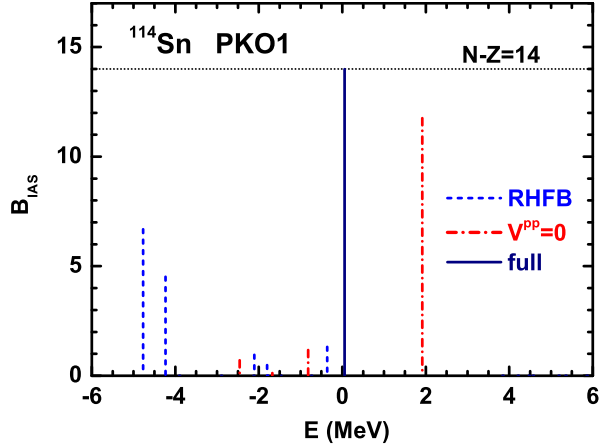}
\caption{(Color online) Transition probabilities for the IAS in $^{114}$Sn. The calculations are performed by the RHFB+QRPA approach with PKO1, while the Coulomb interaction is switched off. The horizontal dotted line denotes the $N-Z$ sum rule. For comparison, the unperturbed result (labelled by RHFB) and the calculation without the pp residual interaction ($V^{pp}=0$) are shown by the dashed and dash-dotted lines, respectively.} \label{Fig:IASCheck}
\end{figure}
As a first test of the present QRPA model, we perform the so-call IAS check to verify the model self-consistency. If the Coulomb interaction is switched off, the nuclear Hamiltonian would commute with the isospin lowering $T_-$ and raising $T_+$ operators and then the IAS should be degenerate with its isobaric multiplet partners. This degeneracy is broken by the mean-field approximation, while it can be restored by the self-consistent RPA calculation \cite{Engelbrecht1970PRL}. Taking the IAS in $^{114}$Sn as an example, the corresponding transition probabilities are shown in Fig.~\ref{Fig:IASCheck}, which are calculated by the RHFB+QRPA approach without the Coulomb interaction. It is found that the unperturbed excitations mainly locate between $E=-5$ and $-4$~MeV, which indicates the isospin symmetry breaking in the RHFB theory. By including the ph residual interactions in the QRPA approach, the transition energy with the largest strength increases to $E=1.9$~MeV, while it still remarkably departs from zero. Furthermore, when the pp residual interactions are included, the energy of IAS goes to $0.05$~MeV and it also exhausts $99.94\%$ of the $N-Z$ sum rule. This indicates the self-consistency is well preserved in the present RHFB+QRPA approach only when the ph and pp residual interactions are both taken into account in the QRPA calculations.

\begin{figure}
\includegraphics[width=8cm]{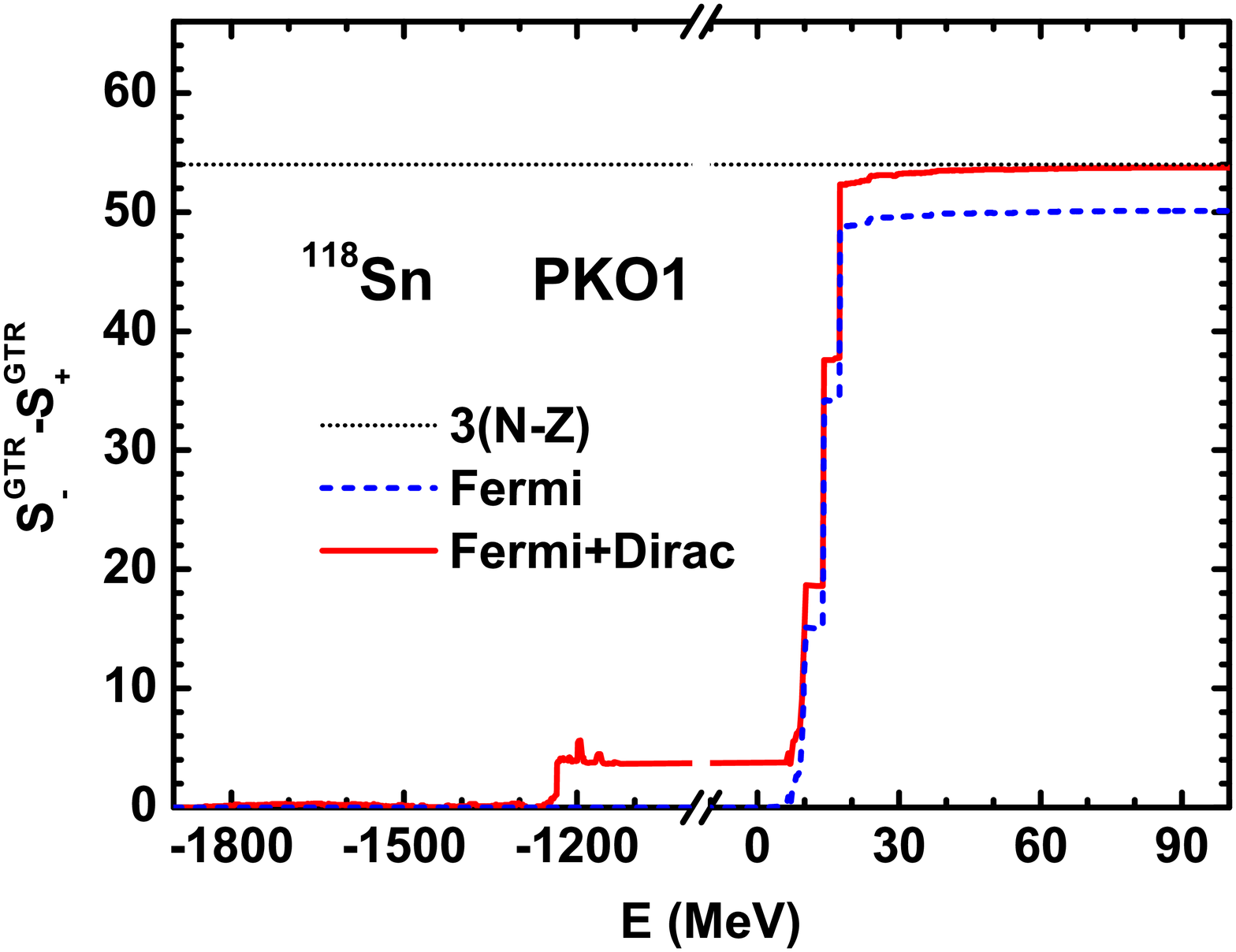}
\caption{(Color online) Running sum of the GT transition probabilities for $^{118}$Sn calculated by the RHFB+QRPA approach with PKO1. The dashed line shows the QRPA calculation with only the ph configurations from the Fermi states. The solid line corresponds to the calculation further including the configurations from the occupied Fermi states and the unoccupied Dirac states. The horizontal dotted line corresponds to the value $3(N-Z)$ of the Ikeda sum rule.} \label{Fig:FigGTRSumRuleSn118}
\end{figure}
As a step further, the sum rule of GT transition probabilities is employed to check the QRPA model. Figure~\ref{Fig:FigGTRSumRuleSn118} presents the running sum of the GT transition probabilities by taking $^{118}$Sn as an example, which is defined to be
\begin{equation}\label{Eq:SumRule}
      (S_{-}^{\textrm{GTR}} - S_{+}^{\textrm{GTR}})_E
    = \sum_{\Omega_\nu<E} (B_{\nu}^{-} - B_{\nu}^{+}),
\end{equation}
where $\Omega_\nu$ represent the GT transition energies and $B_\nu^{\pm}$ are the corresponding transition probabilities in the $T_{\pm}$ channels. When the complete set of states is included, Eq.~(\ref{Eq:SumRule}) gives the value $3(N-Z)$ of the Ikeda sum rule \cite{Ikeda1963PL}.
In the relativistic framework, it has been found that the total GT strength in the nucleon sector is reduced by about $12\%$ in nuclear matter \cite{Kurasawa2003PRL} and by $6\sim7\%$ in finite nuclei \cite{Ma2004EPJA, Liang2008PRL} when compared to the Ikeda sum rule, if the effects related to the Dirac sea are neglected. The dashed line in Fig.~\ref{Fig:FigGTRSumRuleSn118} presents the running sum of the GT transition probabilities calculated with only the ph configurations from the Fermi states. The value of $(S_{-}^{\textrm{GTR}} - S_{+}^{\textrm{GTR}})$ only goes to about $50$ even the sum is extended up to $E=100$~MeV, which is about $7\%$ less than the Ikeda sum rule. When the ph configurations from the occupied Fermi states and the unoccupied Dirac states are further included, they contribute about $4$ to the sum rule even the sum only goes to $E=-1000$~MeV, and this value just compensates the above missing part. This confirms that the total sum rule $3(N-Z)$ is exhausted only when the configurations from the occupied Fermi states and the unoccupied Dirac states are included. Therefore, all the following calculations strictly include these configurations.

\begin{figure}
\includegraphics[width=8.5cm]{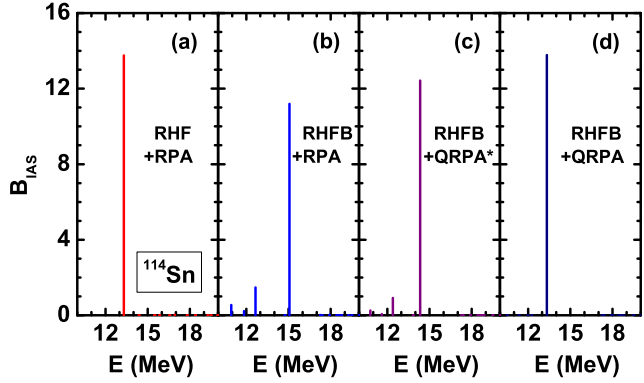}
\caption{(Color online) Transition probabilities for the IAS in $^{114}$Sn calculated with PKO1.
The RHF+RPA, RHFB+RPA, RHFB+QRPA*, and RHFB+QRPA calculations are shown in panels (a), (b), (c), and (d), respectively. See the text for details.}
\label{Fig:FigIASDetailSS}
\end{figure}

The IAS is the simplest but important charge-exchange excitation mode and it has been observed in experiments with a single peak with a narrow width \cite{Pham1995PRC}. It has been found that the consistent treatment of pairing correlations in QRPA calculations plays an essential role in concentrating the IAS in a single peak \cite{Paar2004PRC, Fracasso2005PRC}. In order to investigate such a fact in the RHFB+QRPA approach, Fig.~\ref{Fig:FigIASDetailSS} gives the calculated transition probabilities for the IAS in $^{114}$Sn.

In the panel (a) of Fig.~\ref{Fig:FigIASDetailSS}, the results calculated without any pairing interaction are shown and a single peak is observed. In a sense, the treatment of pairing is consistent here because it is not included in both the ground-state and IAS calculations, but the pairing correlations are essential for open-shell nuclei. The pairing is then included in the RHFB calculation for the ground-state properties, while the pp residual interaction is excluded in the QRPA calculation, which is shown in the panel (b) of Fig.~\ref{Fig:FigIASDetailSS}. It is found that the calculated transition probabilities become fragmented, inconsistent with the experimentally observed single narrow resonance. In addition, the main peak is shifted to higher excitation energy. Furthermore, the direct part of the pp residual interaction is included in the QRPA calculation, and the corresponding results are shown in the panel (c) of Fig.~\ref{Fig:FigIASDetailSS}. The fragmentation of IAS still exists although it has been partially eliminated. In the panel (d) of Fig.~\ref{Fig:FigIASDetailSS}, the fully self-consistent RHFB+QRPA calculation is presented. The IAS is again collected in a single peak, which can exhaust $98\%$ of the $N-Z$ sum rule. Therefore, the consistent treatment of pairing correlations in the QRPA calculation is essential to concentrate the IAS in a single peak, and hence the pp residual interaction has to be incorporated for better understanding the IAS transitions of open-shell nuclei.

\begin{figure}
\includegraphics[width=8cm]{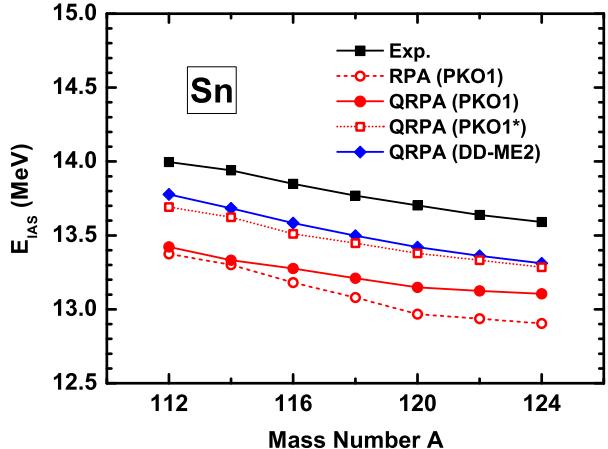}
\caption{(Color online) IAS excitation energies of the even-even Sn isotopes.
The experimental data \cite{Pham1995PRC} are denoted by the filled squares.
The self-consistent RHF+RPA and RHFB+QRPA calculations with PKO1 are shown by the open and filled circles, respectively, while the self-consistent RHB+QRPA calculations with DD-ME2 are shown by the filled diamonds. For comparison, the results obtained with RHFB+QRPA approach with PKO1 but excluding the Coulomb exchange term are denoted by the open squares.} \label{Fig:FigEIAS2A}
\end{figure}

The IAS excitation energies of the even-even Sn isotopes are shown in Fig.~\ref{Fig:FigEIAS2A}.
To investigate the influence of pairing interaction and exchange terms of mean fields, the calculations with the self-consistent RHF+RPA and RHB+QRPA approaches are also shown in addition to the results from the self-consistent RHFB+QRPA calculations. Comparing the results of the self-consistent RHF+RPA and RHFB+QRPA calculations, it is found that the inclusion of $T=1$ pairing interactions can slightly increase the calculated IAS excitation energies. Moreover, it is found that the IAS excitation energies calculated with the RHFB+QRPA and RHB+QRPA approaches are about $300$ and $600$~keV lower than the experimental data.

Since the nonzero IAS excitation energy originates from the existence of the Coulomb field, the different treatments of the Coulomb field would play an important role in understanding this systematic discrepancy between RHFB+QRPA and RHB+QRPA. To verify this argument, we further perform the self-consistent RHFB+QRPA calculations while the Coulomb exchange term is switched off from the beginning.
The corresponding results are shown by the open squares in Fig.~\ref{Fig:FigEIAS2A}.
It is seen that these results are almost the same as those of the RHB+QRPA calculations, so the Coulomb exchange term is responsible for the difference between the IAS excitation energies with the RHFB+QRPA and RHB+QRPA approaches, and the proper treatment of the Coulomb field is important to predict the IAS excitation energies.

\begin{figure}
\includegraphics[width=8cm]{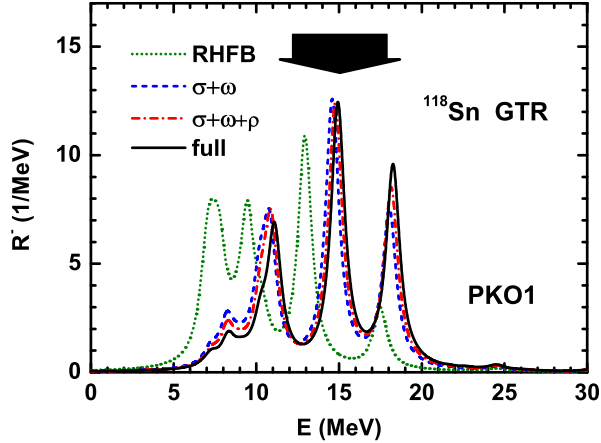}
\caption{(Color online) GT strength distribution in $^{118}$Sn calculated by the RHFB+QRPA approach with PKO1. The unperturbed (labelled by RHFB) strength, the calculation with only ph residual interactions of $\sigma$ and $\omega$ fields, and that with only ph residual interactions of $\sigma$, $\omega$, and $\rho$ fields (excluding $\pi$ field) are shown by the dotted, dashed, and dash-dotted lines, respectively. The experimental data \cite{Pham1995PRC} is shown with an arrow, whose width illustrates the width of the resonance.} \label{Fig:FigGTRDetailPKO1Sn118}
\end{figure}

The GTR is another important mode of charge-exchange excitation and it plays an important role in understanding many nuclear processes in nucleosynthesis, such as nuclear $\beta$ decay and electron-capture process. It has been found that the GTR in the doubly magic nuclei $^{48}$Ca, $^{90}$Zr, and $^{208}$Pb are well reproduced based on the RHF+RPA approach without any readjustment of the ph residual interaction \cite{Liang2008PRL}. In this work, we will check whether such self-consistence is kept even for the open-shell nuclei. In Fig.~\ref{Fig:FigGTRDetailPKO1Sn118}, the GT strength distribution in $^{118}$Sn calculated by the self-consistent RHFB+QRPA approach is shown. It is compared with the unperturbed case, the calculation with only ph residual interactions of $\sigma$ and $\omega$ fields, and that with only ph residual interactions of $\sigma$, $\omega$, and $\rho$ fields. It is clear that the $\sigma$ and $\omega$ mesons play the essential role via the exchange terms, while the $\rho$ and $\pi$ mesons only play a minor role. Similar to the case in the doubly magic nuclei, the experimental excitation energy of the main peak of GTR in open-shell nuclei is also well reproduced by the RHFB+QRPA approach without any readjustment of ph residual interaction.

\begin{figure}
\includegraphics[width=8cm]{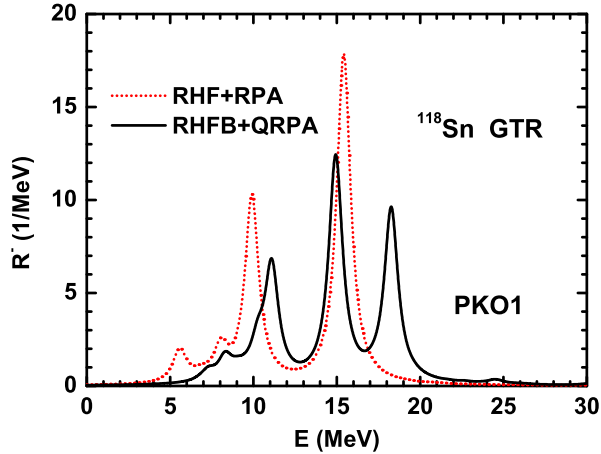}
\caption{(Color online) GT strength distribution in $^{118}$Sn calculated by the RHF+RPA (dotted line) and RHFB+QRPA (solid line) approaches with PKO1.} \label{Fig:FigSn118GTR2Teq1}
\end{figure}

\begin{table}
\begin{center}
\caption{Main neutron-to-proton (Q)RPA amplitudes ($X_{ph}^2-Y_{ph}^2>1\%$) for different GT excitations in $^{118}$Sn calculated by the RHF+RPA and RHFB+QRPA approaches. Excitation energies are in unit of MeV.} \label{Tab:Teq1}
\begin{ruledtabular}
\begin{tabular}{crrcrrr}
Configurations                          &\multicolumn{2}{c}{RHF+RPA} &~~  &\multicolumn{3}{c}{RHFB+QRPA} \\
\cline{2-7}
                                        &$E$=9.9   &15.4             &~~  &11.1      &14.9     &18.3     \\
\hline
$\nu 1g_{ 9/2} \rightarrow \pi 1g_{ 7/2}$   & 5.0\%    &90.5\%           &~~  & 6.4\%    &82.5\%   & 8.4\%   \\
$\nu 1g_{ 7/2} \rightarrow \pi 1g_{ 7/2}$   &10.4\%    & 1.3\%           &~~  & 4.8\%    & 1.1\%   &         \\
$\nu 1g_{ 7/2} \rightarrow \pi 2d_{ 5/2}$   & 2.5\%    &                 &~~  & 2.8\%    &         &         \\
$\nu 2d_{ 5/2} \rightarrow \pi 2d_{ 5/2}$   &12.5\%    &                 &~~  & 6.5\%    &         &         \\
$\nu 2d_{ 5/2} \rightarrow \pi 2d_{ 3/2}$   &57.7\%    & 2.3\%           &~~  &16.7\%    & 2.1\%   &         \\
$\nu 2d_{ 3/2} \rightarrow \pi 2d_{ 5/2}$   & 3.3\%    &                 &~~  & 1.3\%    &         &         \\
$\nu 2d_{ 3/2} \rightarrow \pi 2d_{ 3/2}$   & 6.0\%    &                 &~~  & 3.9\%    &         &         \\
$\nu 2d_{ 3/2} \rightarrow \pi 3s_{ 1/2}$   & 1.7\%    &                 &~~  & 1.2\%    &         &         \\
$\nu 2d_{ 3/2} \rightarrow \pi 3d_{ 5/2}$   &          & 1.5\%           &~~  &          &         &         \\
$\nu 2d_{ 3/2} \rightarrow \pi 3d_{ 3/2}$   &          & 1.7\%           &~~  &          &         &         \\
$\nu 3s_{ 1/2} \rightarrow \pi 3s_{ 1/2}$   &          &                 &~~  & 5.7\%    &         &         \\
$\nu 1h_{11/2} \rightarrow \pi 1h_{11/2}$   &          &                 &~~  &48.0\%    & 1.1\%   &         \\
$\nu 1h_{11/2} \rightarrow \pi 1h_{ 9/2}$   &          &                 &~~  &          &10.1\%   &88.1\%   \\
\end{tabular}
\end{ruledtabular}
\end{center}
\end{table}

Comparing with the doubly magic nuclei, pairing interaction is essential to describe the properties of open-shell nuclei. Figure~\ref{Fig:FigSn118GTR2Teq1} presents the effect of the isovector $T=1$ pairing interaction on the GT strength distribution in $^{118}$Sn.
It is seen that the inclusion of $T=1$ pairing increases the GT energies for transitions below $12$~MeV. For the main peak of GTR, the inclusion of $T=1$ pairing results in the splitting of transition, and the centroid energy in the energy region $12\sim22$~MeV is also increased from $15.4$ to $16.4$~MeV. To understand this GT strength splitting, the main neutron-to-proton (Q)RPA amplitudes ($X_{ph}^2-Y_{ph}^2>1\%$) for different GT excitations in $^{118}$Sn calculated without and with the $T=1$ pairing interaction are given in Table~\ref{Tab:Teq1}. Due to the pairing correlation, the neutrons are scattered to higher levels in $N=50\sim82$ shell, and hence occupy the $h_{11/2}$ level. Therefore, a transition dominated by the new configuration $\nu 1h_{11/2} \rightarrow \pi 1h_{9/2} $ appears and meanwhile the transition at $E\approx 15$~MeV is mixed with new configurations from $\nu 1h_{11/2}$. In addition, the transition at $E=9.9$ MeV is also mixed with a new configuration from $\nu 1h_{11/2}$, whose QRPA amplitude even reaches $50\%$.

\begin{figure}
\includegraphics[width=8cm]{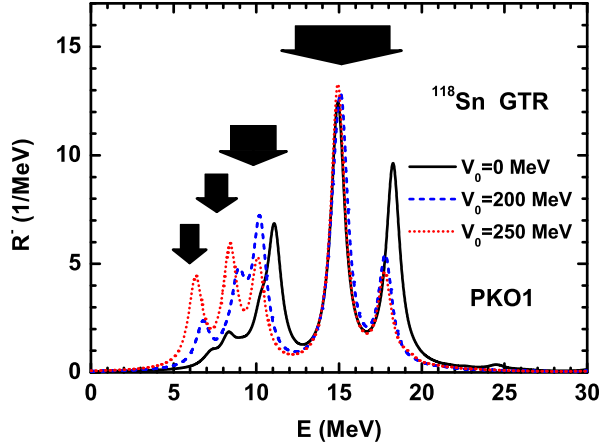}
\caption{(Color online) GT strength distribution in $^{118}$Sn calculated by the RHFB+QRPA approach with PKO1 for different values of $V_0$. The experimental data \cite{Pham1995PRC} are shown with arrows, whose widths illustrate the widths of the corresponding resonances.}
\label{Fig:FigGTR2V0}
\end{figure}

\begin{table}
\begin{center}
\caption{Main neutron-to-proton QRPA amplitudes ($X_{ph}^2-Y_{ph}^2>1\%$) for different GT excitations in $^{118}$Sn calculated by including the $T=0$ pairing interaction with $V_0=250$~MeV. Excitation energies are in unit of MeV.} \label{Tab:Teq0}
\begin{ruledtabular}
\begin{tabular}{ccrcrcr}
Configurations                          &~~  &$E$=10.1  &~   &14.9     &~   &17.8   \\
\hline
$\nu 1g_{ 9/2} \rightarrow \pi 1g_{ 7/2}$   &~~  & 2.7\%    &~   &89.6\%   &~   & 4.6\% \\
$\nu 1g_{ 7/2} \rightarrow \pi 1g_{ 7/2}$   &~~  & 2.8\%    &~   &         &~   &       \\
$\nu 1g_{ 7/2} \rightarrow \pi 2d_{ 5/2}$   &~~  & 1.9\%    &~   &         &~   &       \\
$\nu 2d_{ 5/2} \rightarrow \pi 2d_{ 5/2}$   &~~  & 3.7\%    &~   &         &~   &       \\
$\nu 2d_{ 5/2} \rightarrow \pi 2d_{ 3/2}$   &~~  &81.9\%    &~   & 1.6\%   &~   &       \\
$\nu 2d_{ 5/2} \rightarrow \pi 3d_{ 5/2}$   &~~  &          &~   &         &~   & 4.7\% \\
$\nu 2d_{ 5/2} \rightarrow \pi 3d_{ 3/2}$   &~~  &          &~   &         &~   & 8.6\% \\
$\nu 1h_{11/2} \rightarrow \pi 1h_{11/2}$   &~~  & 2.4\%    &~   &         &~   & 4.5\% \\
$\nu 1h_{11/2} \rightarrow \pi 1h_{ 9/2}$   &~~  &          &~   & 2.7\%   &~   &50.2\% \\
$\nu 1h_{ 9/2} \rightarrow \pi 1h_{11/2}$   &~~  &          &~   & 3.2\%   &~   &29.7\% \\
\end{tabular}
\end{ruledtabular}
\end{center}
\end{table}

In addition to the isovector $T=1$ pairing interaction, the isoscalar $T=0$ pairing interaction also plays an important role in describing the GTR \cite{Engel1999PRC, Paar2004PRC}.
Figure~\ref{Fig:FigGTR2V0} shows the effects of $T=0$ pairing interaction on the GT strength distribution in $^{118}$Sn, where $V_0$ is the strength of the $T=0$ pairing interaction.
Clearly, the excitation energy of the main peak is less affected by the $T=0$ pairing.
However, the $T=0$ pairing interaction reduces the excitation energies and transition strengths in the energy region higher than the main peak, and hence reduces the splitting of GTR in the energy region $12\sim22$~MeV. In the energy region lower than the main peak, the $T=0$ pairing interaction also reduces the excitation energies while it increases the transition strengths. From the QRPA amplitudes for the RHFB+QRPA calculations shown in Table~\ref{Tab:Teq1}, it is known that the main peak at $14.9$~MeV is dominated by the configuration $\nu 1g_{9/2} \rightarrow \pi 1g_{7/2}$, which is almost a pure ph configuration with occupation probabilities $v^2(\nu 1g_{9/2})=0.99$ and $v^2(\pi 1g_{7/2})=0.00$. Therefore, the effect of $T=0$ pairing interaction on the main peak is relatively small. However, the peak at $18.3$~MeV is dominated by the configuration $\nu 1h_{11/2}\rightarrow \pi 1h_{9/2}$, which is more like a pp configuration with occupation probabilities $v^2(\nu 1h_{11/2})=0.21$ and $v^2(\pi 1h_{9/2})=0.00$, and thus the attractive $T=0$ pairing interaction reduces its excitation energy. For the peak at $11.1$~MeV, its main configuration is $\nu 1h_{11/2}\rightarrow \pi 1h_{11/2}$, so the $T=0$ pairing interaction also has an important effect on this transition. For comparison, the main QRPA amplitudes ($X_{ph}^2-Y_{ph}^2>1\%$) for these three GT transitions calculated by including the $T=0$ pairing interaction with $V_0=250$~MeV are given in Table~\ref{Tab:Teq0}. Clearly, the main QRPA amplitudes are remarkably affected by the $T=0$ pairing interaction, especially for those transitions dominated by the pp-type configurations.

\begin{figure}
\includegraphics[width=7.5cm]{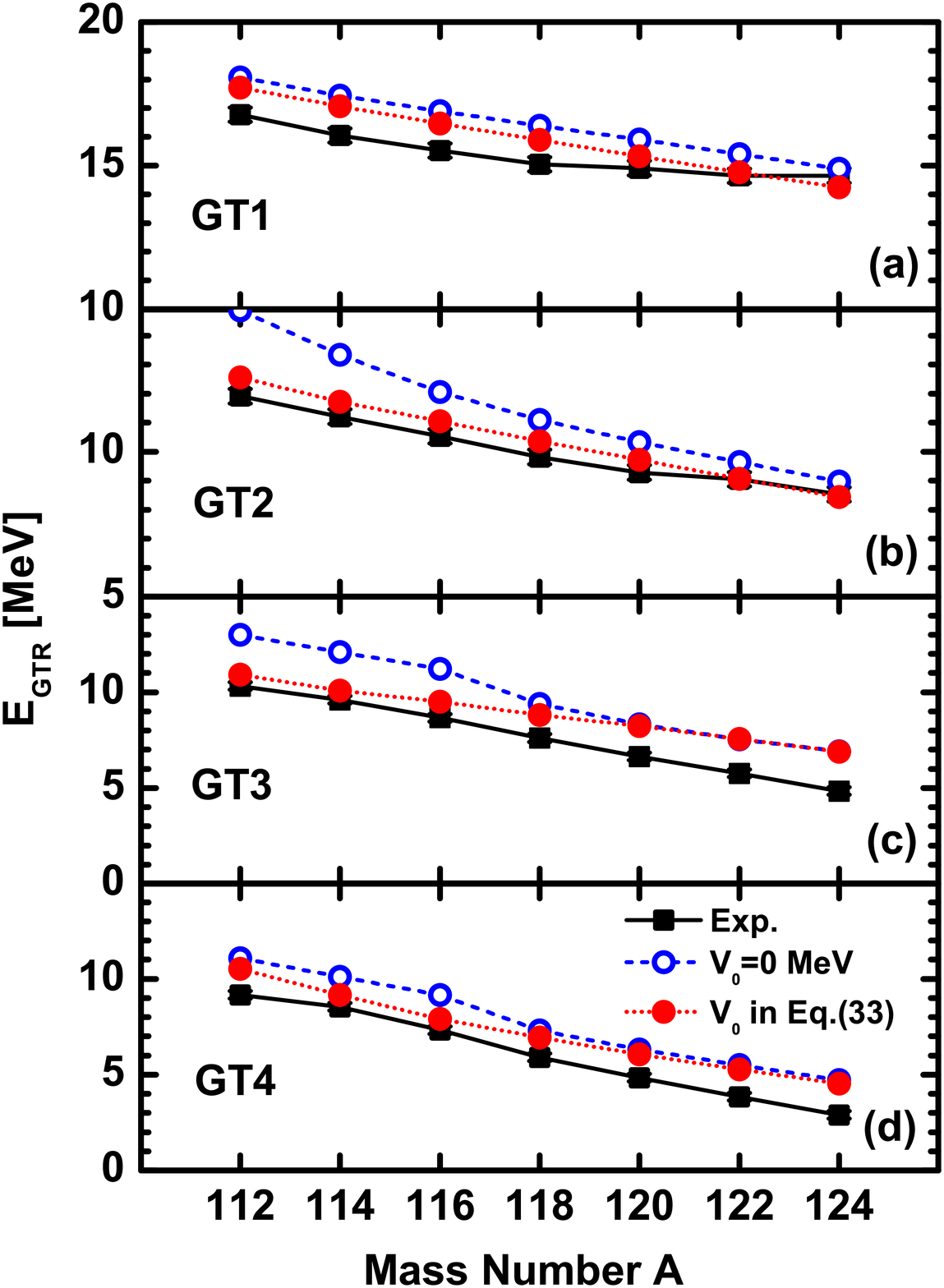}
\caption{(Color online) GT excitation energies of the even-even Sn isotopes. The RHFB+QRPA calculations without and with the $T=0$ pairing in Eq.~(\ref{Eq:V0NSubZ}) are shown by the open and filled circles, respectively. The experimental values in Ref.~\cite{Pham1995PRC} are denoted by the filled squares.} \label{Fig:EGTR12342A}
\end{figure}

For comparison, the experimental GT excitation energies and widths in $^{118}$Sn are also shown in Fig.~\ref{Fig:FigGTR2V0}, which are named to be GT1, GT2, GT3, and GT4 as the decrease of their GT energies similar to Ref.~\cite{Pham1995PRC}. The two peaks in the energy region $12\sim22$~MeV correspond to the GT1, while the predicted splitting of the GTR could not be observed, since the total width of the main resonance is of about $6$~MeV \cite{Pham1995PRC} exceeding the predicted energy splitting. Clearly, the inclusion of $T=0$ pairing interaction improves the theoretical description of low-lying GT transitions. Then the GT2, GT3, and GT4 in $^{118}$Sn are well predicted by the RHFB+QRPA approach.

The strength $V_0$ of $T=0$ pairing interaction is usually determined by fitting to the measured nuclear $\beta$-decay half-lives. A recent study based on the RHFB+QRPA approach found that an isospin-dependent $V_0$ can provide a good description of nuclear $\beta$-decay half-lives in the region of $20\leqslant Z\leqslant 50$ \cite{Niu2013PLB}. With this isospin-dependent $V_0$ shown in Eq.~(\ref{Eq:V0NSubZ}), the calculated centroid energies for the GT1, GT2, GT3, and GT4 of the even-even Sn isotopes are shown in Fig.~\ref{Fig:EGTR12342A}. Without the $T=0$ pairing interaction, the GT excitation energies are systematically higher than the experimental data. The $T=0$ pairing interaction can reduce the GT excitation energies and the agreements with the experimental data are improved systematically. In addition, it is found that the influence of $T=0$ pairing on the excitation energies of GT2, GT3, and GT4 decreases as the neutron number increases. This can be understood from the fact that the pairing effects become weaker and weaker when approaching the closed shell $N=82$.

\subsection{Nuclear $\beta$ decays}

The GT transitions are the dominant transitions in nuclear $\beta$ decays. With the transition energies and strengths of GT excitations, nuclear $\beta$-decay half-lives can be calculated by using Eq.~(\ref{Eq:BetaDecayRate}).

\begin{figure}
\includegraphics[width=7cm]{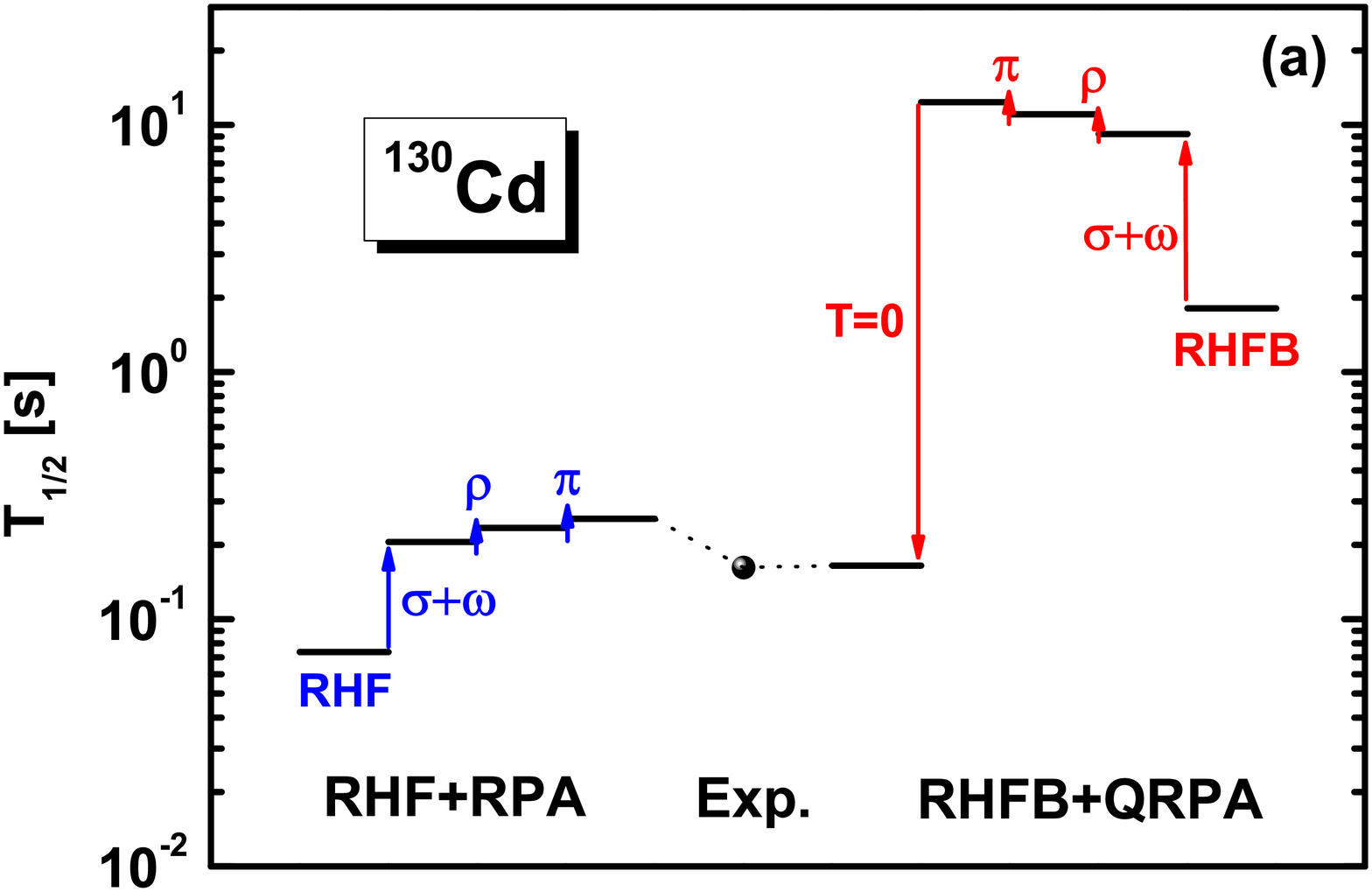}
\includegraphics[width=7cm]{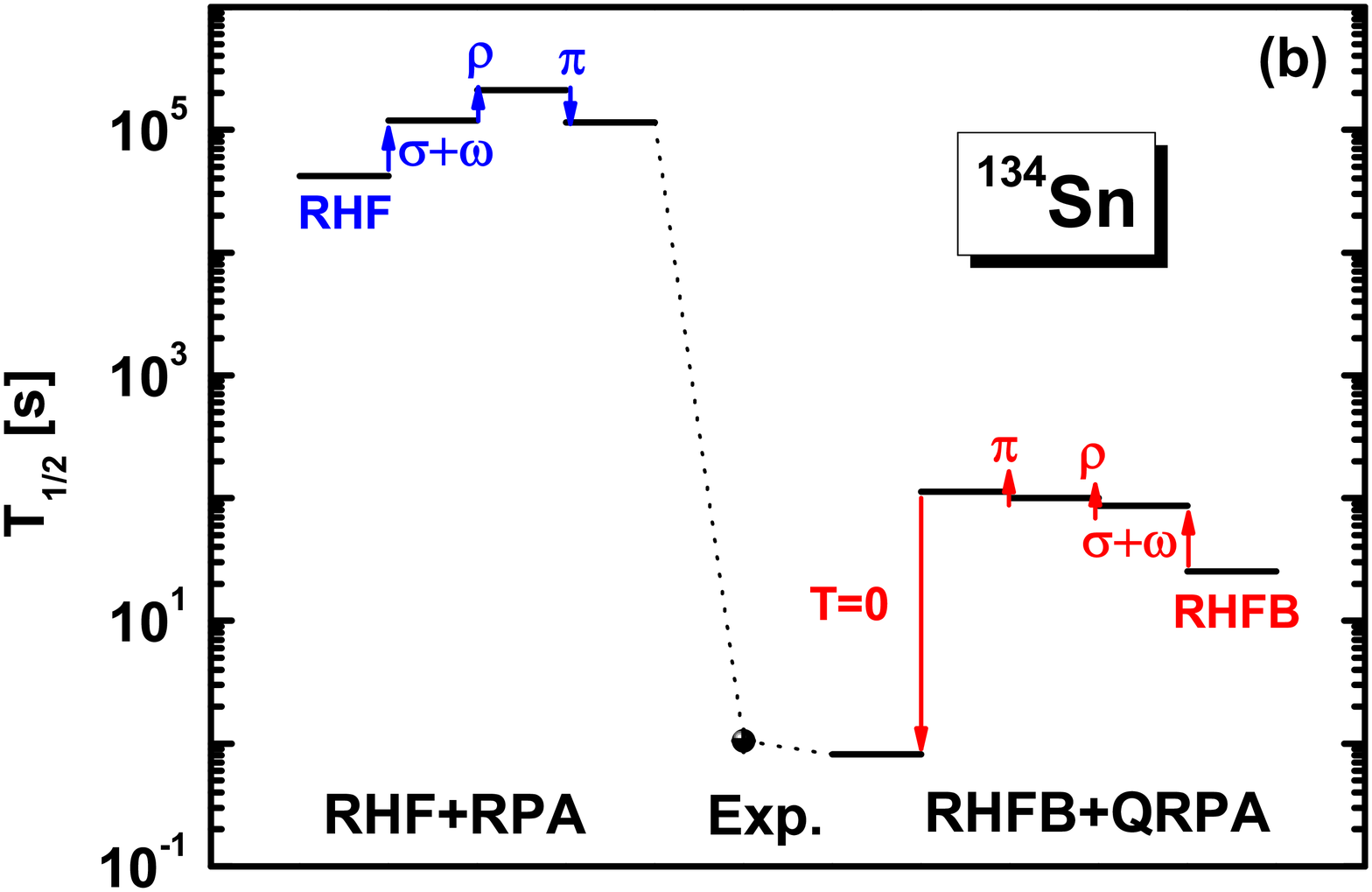}
\caption{(Color online) Nuclear $\beta$-decay half-lives of $^{130}$Cd and $^{134}$Sn calculated by the RHF+RPA and RHFB+QRPA approaches with PKO1. The results based on the (Q)RPA calculations without any residual interactions [labelled to be RHF(B)] and the calculations gradually including the residual interactions of $\sigma$ and $\omega$ fields, $\rho$ field, $\pi$ field, and $T=0$ pairing are presented. The experimental half-lives are shown for comparison.}
\label{Fig:DetailBetaParing}
\end{figure}

First, let us investigate the effects of various residual interactions in the RHFB+QRPA calculations on predicting nuclear $\beta$-decay half-lives, which are shown in Fig.~\ref{Fig:DetailBetaParing} by taking $^{130}$Cd and $^{134}$Sn as examples. Comparing the results between RHF and RHFB without any residual interaction, it is found that the $T=1$ pairing plays an important role in predicting nuclear $\beta$-decay half-lives, which are increased by about an order of magnitude for $^{130}$Cd while reduced by three orders of magnitude for $^{134}$Sn. Furthermore, the ph residual interactions from $\sigma$ and $\omega$ fields, $\rho$ field, and $\pi$ field are gradually included. It is found that the $\sigma$ and $\omega$ fields play an essential role comparing with the $\rho$ and $\pi$ fields. In total, the ph residual interactions increase the calculated $\beta$-decay half-lives. However, the RHFB+QRPA calculations with all ph residual interactions overestimate the nuclear $\beta$-decay half-lives by about two orders of magnitude. From Fig.~\ref{Fig:FigGTR2V0}, it is known that the attractive $T =0$ pairing interaction works to reduce the transition energies---that increase the phase volume $f(Z, E_m)$ in Eq.~(\ref{Eq:PhaseVol})---and increase transition strengths, therefore, the inclusion of $T = 0$ pairing in general reduces the $\beta$-decay half-lives. With the strengths of $V_0$ proposed in Ref.~\cite{Niu2013PLB}, the RHFB+QRPA calculations well reproduce the experimental nuclear $\beta$-decay half-lives of $^{130}$Cd and $^{134}$Sn.

\begin{figure}
\includegraphics[width=8cm]{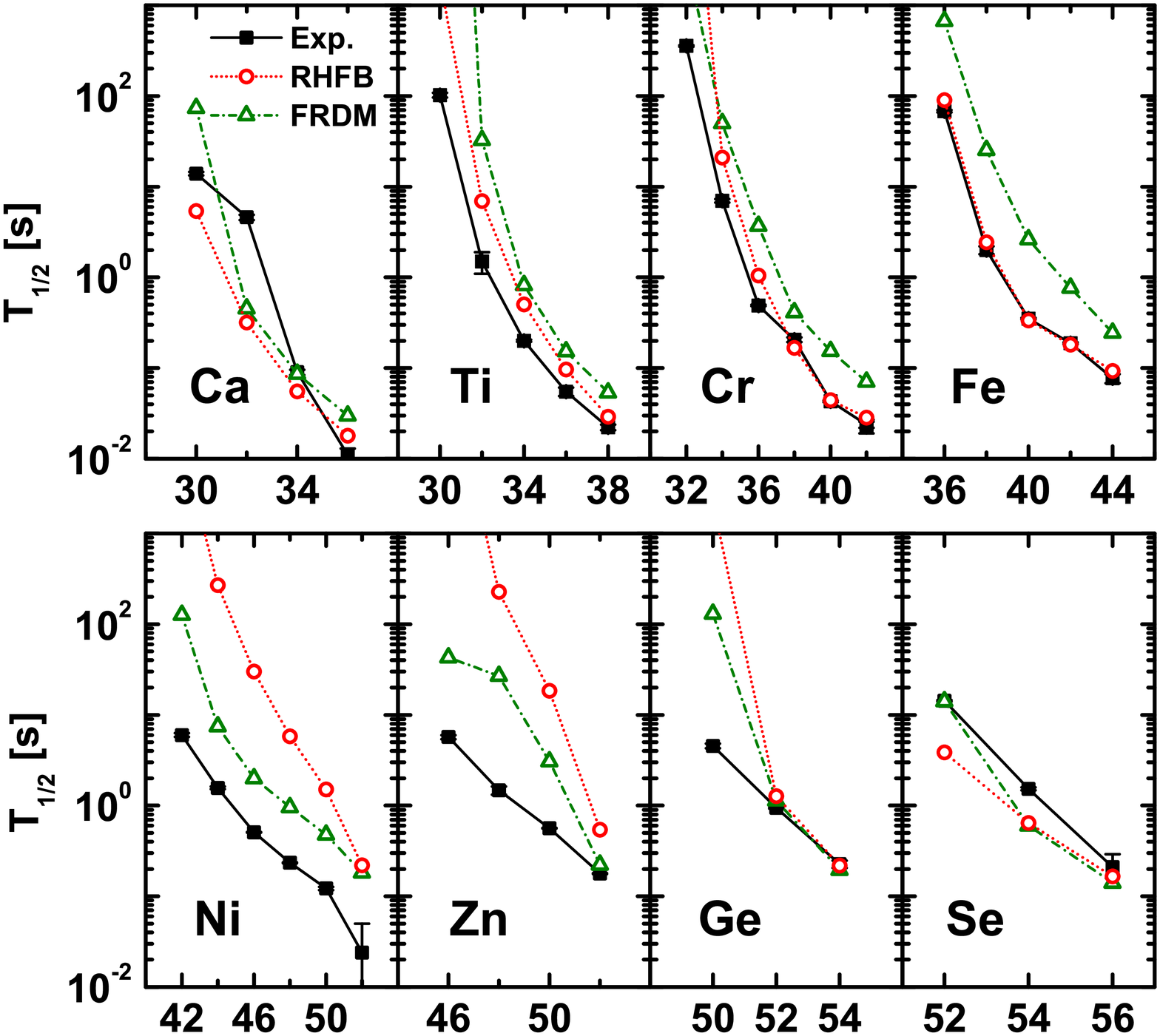}
\includegraphics[width=8cm]{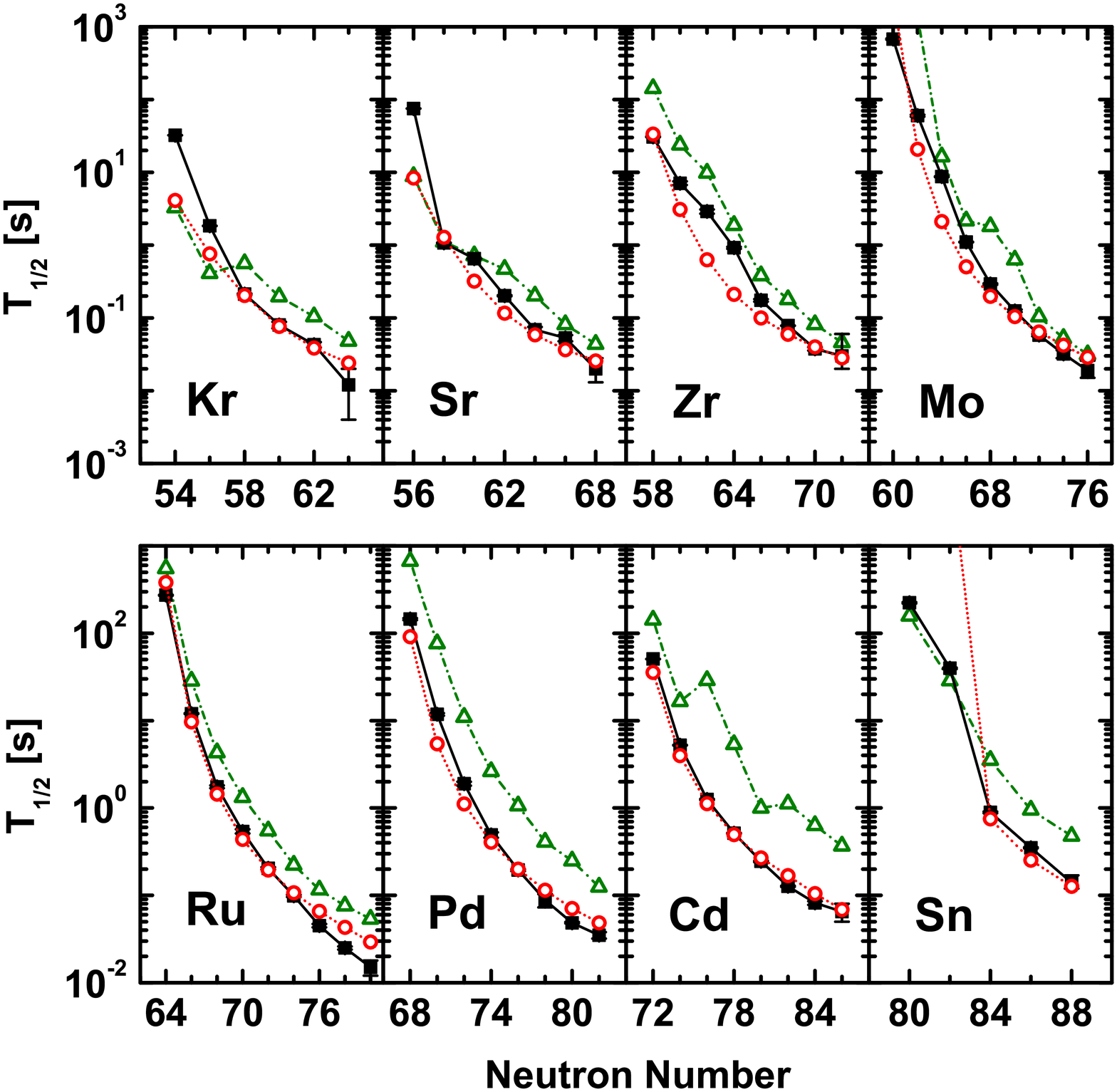}
\caption{(Color online) $\beta$-decay half-lives of even-even nuclei with $20\leqslant Z\leqslant 50$ calculated by the RHFB+QRPA approach with PKO1 (open circles). For comparison, the theoretical results obtained in the FRDM+QRPA calculations and the experimental values in NUBASE2012 \cite{Audi2012CPC} updated with new data in Refs.~\cite{Lorusso2015PRL, Xu2014PRL, Mazzocchi2013PRC} are shown by the filled squares and open triangles, respectively.} \label{Fig:BetamFitFeCdV0PExp}
\end{figure}

Furthermore, the $\beta$-decay half-lives of even-even nuclei with $20\leqslant Z\leqslant 50$ are calculated by the RHFB+QRPA approach with the isospin-dependent $V_0$ \cite{Niu2013PLB}. The corresponding results are shown in Fig.~\ref{Fig:BetamFitFeCdV0PExp} together with the results by the FRDM+QRPA approach and the experimental data. It is seen that the FRDM+QRPA approach almost systematically overestimates the experimental half-lives in this region of nuclear chart. It has been pointed out that the overestimation of half-lives in the FRDM+QRPA approach can be attributed partly to the neglect of the $T=0$ pairing \cite{Engel1999PRC, Niu2013PLB}. Comparing with the FRDM+QRPA results, the RHFB+QRPA approach well reproduces the experimental half-lives of these neutron-rich nuclei, except for the Ni, Zn, Ge, and Sn isotopes with neutron number smaller than the corresponding neutron shell, i.e., $N=50$ for the Ni, Zn, and Ge isotopes and $N=82$ for the Sn isotopes. The overestimation of these nuclear half-lives can be understood from the main configurations of the transitions dominating their $\beta$ decays. These main configurations are generally formed by the neutron levels with higher occupation probabilities and the proton levels with lower occupation probabilities, therefore, the influence of $T=0$ pairing interaction is very small and hence their $\beta$-decay half-lives are overestimated. In fact, this phenomenon is a common problem in the self-consistent relativistic QRPA calculations \cite{Niksic2005PRC, Marketin2007PRC, Niu2013PLB, Wang2016JPG}.

\begin{figure}
\includegraphics[width=8cm]{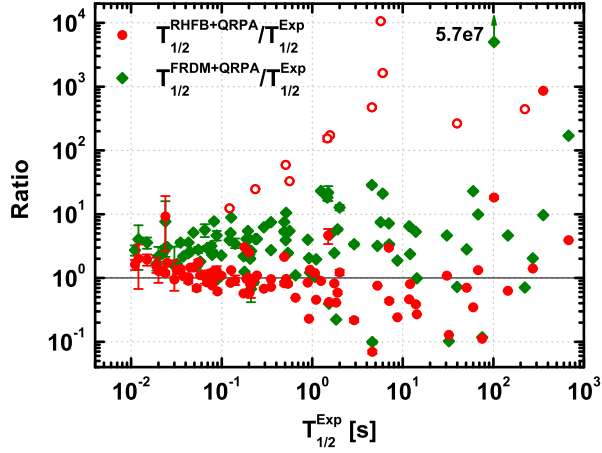}
\caption{(Color online) Ratios of the theoretical $\beta$-decay half-lives to the experimental data as a function of the experimental half-lives for the even-even nuclei with $20\leqslant Z\leqslant 50$.
The ratios corresponding to the RHFB+QRPA approach and the FRDM+QRPA approach are denoted by the circles and squares, respectively. The open circles correspond to those for the Ni, Zn, Ge, and Sn isotopes with neutron number smaller than the corresponding neutron shell.} \label{Fig:Ratio2T12FitFeCdV0}
\end{figure}

To investigate the reliability of theoretical approaches in various half-life regions, Fig.~\ref{Fig:Ratio2T12FitFeCdV0} presents the ratios of the theoretical $\beta$-decay half-lives to the experimental data as a function of the experimental half-lives. As discussed above, the ratios calculated by the RHFB+QRPA approach for the Ni, Zn, Ge, and Sn isotopes with neutron number smaller than the corresponding neutron shell are remarkably larger than those of other nuclei. In general, the half-lives of $T_{1/2}^{\rm Exp}<1$~s are almost completely reproduced, and those of $1~\textrm{s}<T_{1/2}^{\rm Exp}<100$~s are reproduced within an order of magnitude, while the results show relatively larger scattering for the nuclei with $T_{1/2}^{\rm Exp}>100$~s. In other words, the average error in $\beta$-decay half-life description increases as the half-life increases, which is also observed for the results of FRDM+QRPA approach. The long-lived nuclei are more sensitive to small shifts in the positions of the calculated GT transitions, so the half-life calculations are more reliable for nuclei far from stability than those close to $\beta$-stability line, presenting a correlation between the average error and the experimental $\beta$-decay half-life. In addition, the overestimation of $\beta$-decay half-life is also clearly found for the FRDM+QRPA approach.

\begin{figure}
\includegraphics[width=7cm]{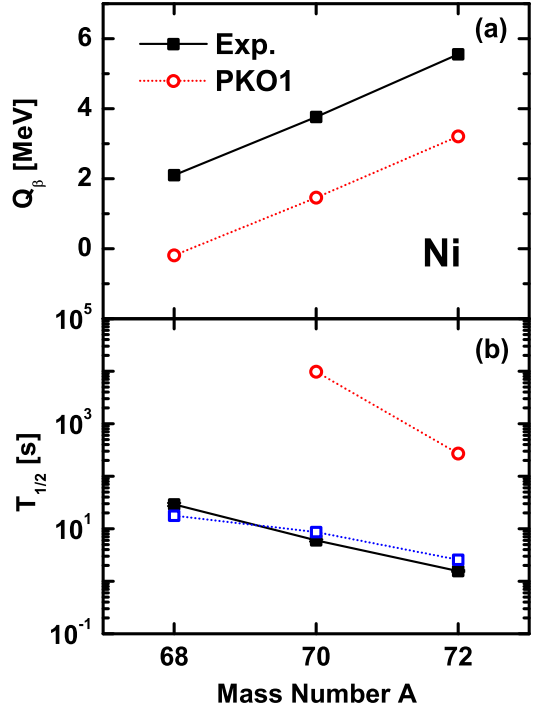}\\
\caption{(Color online) $Q_\beta$ values [panel (a)] and its influence on $\beta$-decay half-lives [panel (b)] of the Ni isotopes. The open circles denote the results calculated with PKO1 based on the RHFB and RHFB+QRPA approaches, respectively. The experimental data are shown with the filled squares. The open squares are the same as the open circles but replacing the calculated $E_M$ by $E_M-\Delta Q_\beta$, where $\Delta Q_\beta$ is the difference of $Q_\beta$ between the experimental value \cite{Wang2012CPC} and RHFB calculation.}\label{Fig:Qbeta}
\end{figure}

The $Q_\beta$ value plays a crucial role in determining nuclear $\beta$-decay half-lives, so its effect may help to improve the description of $\beta$-decay half-lives for the Ni, Zn, Ge, and Sn isotopes with neutron number smaller than the corresponding neutron shell. By taking the Ni isotopes as examples, Fig.~\ref{Fig:Qbeta} presents the $Q_\beta$ values and its influence on $\beta$-decay half-lives. It is clear that the experimental $Q_{\beta}$ values of the Ni isotopes are systematically underestimated by the RHFB theory. To further estimate the influence of $Q_{\beta}$ values on the half-life predictions, the half-lives calculated by merely replacing $E_M$ by $E_M-\Delta Q_\beta$ are shown by the open squares in Fig.~\ref{Fig:Qbeta}. It is striking that the new results are in excellent agreement with the experimental data, which reflects the importance of $Q_{\beta}$ value in half-life calculations.

It should be pointed out that this modification of $E_M$ with $\Delta Q_\beta$ is not a self-consistent prediction for nuclear $\beta$-decay half-lives. Recent self-consistent RPA calculations in the non-relativistic framework found that the inclusion of an attractive tensor force can reduce the calculated half-lives of magic nuclei \cite{Minato2013PRL}. However, new parameters for the tensor force are inevitable. By taking into account the coupling between particles and collective vibrations, the self-consistent RPA plus particle-vibration coupling (PVC) model can well reproduce the half-lives of magic nuclei without any new fitting parameters \cite{NiuYF2015PRL}. In present model, the effects of the tensor force have been indeed involved via the exchange diagrams of meson-nucleon couplings which have been demonstrated to contain the tensor force components~\cite{Jiang2015-1}, whereas the PVC effects are not included yet. Thus, part of these effects in open-shell nuclei may be simulated by the $T=0$ pairing through the enhanced pairing strength. When a self-consistent relativistic QRPA model with all these effects is developed in the future, the $T=0$ pairing strength may need to be readjusted, and this would help to further understand the importance of $T=0$ pairing in the half-life predictions.

\section{Summary and perspectives}\label{Sec:4}

In this work, the self-consistent quasiparticle random-phase approximation model is developed based on the relativistic Hatree-Fock-Bogoliubov theory, and it is then employed to study the nuclear isobaric analog states and Gamov-Teller resonances by taking Sn isotopes as examples. It is found that the particle-particle residual interaction is essential to concentrate the IAS in a single peak for open-shell nuclei and the Coulomb exchange terms are very important to predict the IAS energies. For the GTR, the isoscalar $\sigma$ and $\omega$ mesons play an crucial role in the particle-hole residual interactions via the exchange terms. The isovector pairing can increase the calculated GTR energies and result in new excitations as the pairing scatters nucleons to higher energy levels. The isoscalar pairing has a strong influence on the low-lying tail of the GTR and is necessary to reproduce the experimental GTR energies. With the predicted properties of GT transitions by the QRPA approach, nuclear $\beta$-decay half-lives are studied in the allowed Gamow-Teller approximation. Among the particle-hole residual interactions, $\sigma$ and $\omega$ mesons play an important role in the $\beta$-decay calculations. The pairing interactions in both isovector and isoscalar channels are important to reproduce experimental $\beta$-decay half-lives. With the results predicted by the RHFB+QRPA approach, the $\beta$-decay calculations almost completely reproduce the experimental data for nuclei with $T_{1/2} < 1$~s up to the Sn isotopes. Large discrepancies are found for the Ni, Zn, Ge, and Sn isotopes with neutron number smaller than the corresponding neutron shell, which can be remarkably improved when the theoretical $Q_\beta$ values are replaced by the corresponding experimental data.

The present RHFB+QRPA approach can also be employed to study other nuclear charge-exchange excitations, such as the spin-dipole and spin-quadrupole resonances. The predicted properties of charge-exchange excitations can be further used to calculate other nuclear weak-interaction processes, such as nuclear electron capture and neutrino-nucleus scattering. In addition, the present QRPA approach are formulated with the spherical symmetry, so it is worthwhile to extend the present approach by including deformation degree of freedom in the future for better describing the properties of deformed nuclei.

\section{Acknowledgements}

This work was partly supported by the National Natural Science Foundation of China (Grants No.
11205004, No.11305161, No. 11335002, No. 11375076, and No. 11411130147), the Key Research Foundation of Education Ministry of Anhui Province of China under Grant No. KJ2016A026, the Specialized Research Fund for the Doctoral Program of Higher Education under Grant No. 20130211110005, and the RIKEN iTHES project.



\end{CJK*}
\end{document}